\documentclass [10pt, twocolumn, conference] {IEEEtran}
\usepackage{amssymb}
\usepackage{graphicx}
\usepackage{multirow}
\usepackage{epstopdf}
\usepackage{setspace}
\usepackage{booktabs,caption,fixltx2e}
\usepackage[flushleft]{threeparttable}
\usepackage{algorithm,algpseudocode}
\usepackage{enumitem}
\usepackage{amsmath}
\usepackage{comment}
\usepackage{subcaption}
\usepackage{caption}
\usepackage{balance}
\usepackage{siunitx}

\usepackage{diagbox}
\graphicspath {{./images}}
\algrenewcomment[1]{\bgroup\hfill//~#1\egroup}

\usepackage{color}
\usepackage{romannum}
\usepackage{color,soul}

\begin{document}


\title{Security Vulnerabilities in Quantum Cloud Systems: A Survey on Emerging Threats}

\author{Justin Coupel, Tasnuva Farheen\\Division of Computer Science, Louisiana State University}

\maketitle

\begin{abstract}
Quantum computing is becoming increasingly widespread due to the potential and capabilities to solve complex problems beyond the scope of classical computers. As Quantum Cloud services are adopted by businesses and research groups, they allow for greater progress and application in many fields. However, the inherent vulnerabilities of these environments pose significant security concerns. This survey delivers a comprehensive analysis of the security challenges that emerged in quantum cloud systems, with a distinct focus on multi-tenant vulnerabilities and the classical-quantum interface. Key threats such as crosstalk attacks, quantum-specific side-channel vulnerabilities, and insider threats are all examined, as well as their effects on the confidentiality, integrity, and availability of quantum circuits. The design and implementation of various quantum architectures from quantum cloud providers are also discussed. In addition, this paper delves into emerging quantum security solutions and best practices to mitigate these risks. This survey offers insights into current research gaps and proposes future directions for secure and resilient quantum cloud infrastructures.
\end{abstract}


\section{Introduction}
Quantum computing has emerged as a revolutionary technology capable of solving complex problems beyond the realm of classical computers with its exponentially increased efficiency. There are new advances in research for multiple aspects of the field, such as quantum algorithms, security measures, and vulnerabilities in quantum systems. At the same time, hardware is improving with an increasing number of qubits in modern quantum processors. Many industries are exploring the potential applications of quantum computing in fields such as cryptography, medical research, and artificial intelligence, making it a key player in the future of scientific and technological innovation.

In recent years, the ability to experiment with quantum hardware has become more accessible than ever. Cloud computing has become the backbone of modern digital services, offering scalable computing resources, storage, and specialized platforms to fit the needs of researchers and enterprises. There are many quantum platform providers such as IBM Quantum, Google, D-Wave, IonQ, Rigetti, and others that provide cloud-based access to quantum computers~\cite{nguyen2024quantum}. These platforms enable developers to experiment with quantum processors remotely, reducing the need for costly on-premises infrastructure. Quantum cloud services provide users with various tools, including quantum simulators, software development kits (SDKs), development environments, and algorithm libraries, which facilitate experimentation and innovation~\cite{singh2024survey}.

There are many types of known threats posed by quantum cloud systems, ranging from vulnerabilities in quantum hardware, classical components of these systems, the quantum-classical interface, or, specifically, in multi-tenant environments. The confidentiality, integrity, and availability of quantum cloud systems can all be targeted as a result of these attacks. Threat actors could exploit vulnerabilities to gain unauthorized access, perform side-channel attacks, reduce availability through denial-of-service (DoS) attacks, or manipulate quantum computations. The consequences of attacking these systems could lead to data breaches and intellectual property theft. One threat to the classical-quantum interface is that of insiders who have access to room-temperature electronics. Through side-channel leakage, this attack could potentially decode internal signals~\cite{mustafa2024side}. Thus, it is important to be aware of all the vulnerabilities in quantum cloud systems, especially those that are emerging and receive less coverage, such as multi-tenancy.

Multi-tenancy is a core feature of cloud platforms, allowing several users to simultaneously share the same physical or virtual resources for testing on quantum computers. This design gives cloud providers lower costs for both hardware and software, as well as optimization of performance~\cite{su2015modeling}. Although this widespread access to quantum resources promotes increased research and discoveries, it can have the disadvantage of introducing new problems that are not present in restricted single-user platforms. A bug in software or hardware activated by one user would negatively impact the experience of other users sharing quantum hardware~\cite{karatacs2017multi}. Consequently, a threat actor can introduce many unique security risks to these shared environments, which could affect many unknowing users through various means. 

Weak security in a multi-tenant quantum cloud environment can have severe implications. It includes many threats that single-tenant systems are exposed to, as well as its own problems. Many side-channel attacks can easily predict the circuits of users with increasingly limited knowledge and are challenging to defend under this system. Adversaries can detect crosstalk and timing patterns in these shared systems and utilize them for exploits. An example in which malicious actors can initiate attacks that utilize crosstalk in NISQ computers can result in circuit discovery~\cite{choudhury2024crosstalk}. Therefore, it is critical to ensure that robust security measures are in place to prevent such risks and protect sensitive quantum workloads, especially for research in crucial sectors such as medicine and finance.

Previous research has covered the concept of multi-tenancy in quantum computing and how this system can be exploited~\cite{choudhury2024survey}. There are many different proposed side-channel exploits that utilize various different methods. Research has also touched on threat vectors in the classical quantum interface and other attack types in quantum cloud systems~\cite{mustafa2024side}. Additionally, there are proposals for mitigation techniques that can reduce the risk of data breaches under multi-tenant and single-tenant frameworks. However, there is still a lot of room for more research into the specific challenges and vulnerabilities posed by multi-tenancy, the classical-quantum interface, and the specific solutions to these issues. 

Key gaps that this survey paper addresses include the concise organization of many of the new vulnerabilities and solutions proposed by recent and ongoing research. There is also a lack of visual representations that can break down many complex systems like the architectures of quantum hardware, quantum cloud providers, the classical-quantum interface, and the multi-tenant model. Furthermore, research gaps and challenges are also presented to accomplish in the future, emphasizing vulnerability assessment of the exploits on other quantum hardware and creating secure solutions of the existing attack vectors. On the part of the cloud providers, monitoring and auditing mechanisms for quantum workloads are essential to ensure the integrity and confidentiality of computations. There is also room for more research on quantum-safe cryptographic algorithms that ensure data protection in multi-tenant quantum cloud systems to bolster its security~\cite{hayat2024securing}.

\textbf{Contributions:} The main contributions of this paper are described below:

\begin{enumerate}
    \item  We present a comprehensive yet concise survey focusing on the security threats in quantum cloud systems and their effects on the confidentiality, integrity, and availability of a user's circuit.
    \item We discuss many different attack vectors on these cloud systems, from the classical-quantum interface, single-tenant threats, multi-tenant threats, insider attacks, quantum hardware attacks, and classical component attacks in quantum devices.
    \item We describe the architecture of a quantum computer and its hardware from a high level with aid from visuals.
    \item We discuss many proposed mitigation strategies and secure solutions in terms of their effectiveness in reducing threats and their feasibility.
    \item We highlight different areas that could require further research and proposals on improving the security in quantum cloud systems.
\end{enumerate}


\begin{table*}[ht]
\caption{Summary of Notable Research on Quantum Computing Security and Multi-Tenancy}
\label{fig:RelatedWorks}
\centering
\renewcommand{\arraystretch}{1.2}
\small
\begin{tabular}{|>{\centering\arraybackslash}p{6.6cm}|>{\centering\arraybackslash}p{9.4cm}|}
\hline
\textbf{Notable Research Work} & \textbf{Paper Coverage} \\
\hline
Side-channel Attacks Targeting Classical-Quantum Interface in Quantum Computers~\cite{mustafa2024side} &
Explains many different attack vectors that affect the classical-quantum interface on quantum hardware, mainly focusing on insider threat potential. \\
\hline
Crosstalk-induced Side Channel Threats in Multi-Tenant NISQ Computers~\cite{choudhury2024crosstalk} &
Proposes crosstalk attack on multi-tenant systems to recover circuit information from victims in order to fully reproduce circuits using limited initial knowledge. \\
\hline
Securing the Cloud Infrastructure: Investigating Multi-tenancy Challenges, Modern Solutions and Future Research Opportunities~\cite{choudhury2024survey} &
A broad survey paper that discusses the development of various solutions to combat the many threats multi-tenancy poses. It suggests encryptions, anti-virus, and more. \\
\hline
A reference architecture for quantum computing as a service~\cite{hayat2024securing} &
Describes Quantum Computing as a Service and the design decisions that platform providers make and implement. \\
\hline
Distributed quantum computing: a survey~\cite{ahmad2024reference} &
Explains how distributed quantum computing systems operate in NISQ environments. \\
\hline
Technological diversity of quantum computing providers: a comparative study and a proposal for API Gateway integration~\cite{caleffi2024distributed} &
Extensive research on the common quantum platform providers and comparing them based on performance, hardware used, and pricing. \\
\hline
A survey of side-channel attacks in superconducting quantum computers~\cite{alvarado2024technological} &
Delivers information on many different side-channel exploits for single-tenant and multi-tenant quantum systems on the cloud with crosstalk, time-based attacks, fault injections, and power-based attacks. \\
\hline
Multi-Tenancy in Cloud Computing~\cite{aljahdali2014multi} &
Provides information on multi-tenancy in the classical sphere and its impact overall in cloud computing. It can shed light to similar concepts applied to quantum platforms. \\
\hline
Quantum leak: Timing side-channel attacks on cloud-based quantum services~\cite{lu2024quantum} &
Discusses several timing-based side-channel attacks on the quantum cloud to identify circuit information and unique quantum hardware. \\
\hline
Detecting fraudulent services on quantum cloud platforms via dynamic fingerprinting~\cite{lu2024quantum} &
Proposes the idea that quantum platform providers switch the hardware users connect to without their permission and creates a method to detect these switches. \\
\hline
Enhancing security and privacy in advanced computing systems: A comprehensive analysis~\cite{mendoza2023enhancing} &
Examines several different security and privacy defense mechanisms to improve security and quantum systems such as data at rest encryption. \\
\hline
\end{tabular}
\end{table*}

\section{Related Works}
This section highlights key references and prior research that have contributed to the understanding of quantum computers and the security threats in multi-tenant quantum cloud environments. The table~\ref{fig:RelatedWorks} includes survey papers on quantum cloud security and multi-tenant systems. In addition, this section explores ongoing research on new security threats, exploits, and mitigation strategies for these vulnerabilities. 

\subsection{Quantum Cloud Research}
Here, we examine previous works on cloud-based infrastructures developed by major service providers. There have been works describing the structure of Quantum Computing as a Service (QCaaS), with the design and implementation that quantum platform providers use~\cite{ahmad2024reference}. There has also been progress in surveying distributed quantum computing NISQ systems and how quantum computers operate in this environment~\cite{caleffi2024distributed}. Research on a comparison between the major platform providers including IBM Quantum, Google Quantum AI, Azure Quantum, and Amazon Bracket has also been discussed using the Traveling Salesman Problem~\cite{alvarado2024technological}. 

\subsection{Security Threats Papers}
In this section, we cover many side-channel attacks that affect superconducting quantum computers. The following paper includes some single-tenant and multi-tenant threats ranging from power-based attacks, timing-based attacks, fault injections, and crosstalk exploits~\cite{choudhury2024survey}. There is also research in power-based side-channel vulnerabilities in quantum computer controllers so this topic will not be focused on this paper~\cite{xu2023exploration}. Additionally, there is much coverage on specific quantum hardware issues and how machine learning (ML) can potentially be used to help mitigate these threats~\cite{karimi2021hardware}. The impact of ML on quantum security is heavily researched on with discussions relating to quantum defenses like adversarial training, data privacy, and formal verification methods in this paper~\cite{franco2024predominant}. In addition, the classical-quantum interface is vulnerable to attacks especially by knowledgeable insiders who can analyze information on user circuits through passive monitoring on SFQ chips~\cite{mustafa2024side}. There are also proposed attack vectors of reverse engineering SFQ chips to recover circuits~\cite{kumar2019toward}. Overall, this topic has much less research, so it will be covered more extensively later.

\subsection{Multi-Tenancy Research}
Numerous works have been published on the usage of multi-tenancy in the classical realm, exploring its structure and vulnerabilities~\cite{aljahdali2014multi}. However, research on multi-tenancy in the quantum cloud is much more limited and typically focuses on analyzing the threats that it creates. There have been several proposed attacks against multi-tenant quantum platforms to discover information about the circuits of other users. The crosstalk created by NISQ multi-tenant computers has been shown to help extract unauthorized information on the victim's circuit by determining the number of CNOT gates in a quantum computer~\cite{choudhury2024crosstalk}. This prior paper also introduces a framework for a side-channel attack utilizing this crosstalk in NISQ systems with the aid of a graph-based model. However, crosstalk has been implemented for several side-channel attacks with the ultimate goal of rebuilding a quantum circuit that should have remained confidential. It has been a known exploit for some time, with multiple sources mentioning threats it may pose. There is another side-channel attack that can inflict potentially major disruptions on victim circuits using the SWAP path in active or passive attacks~\cite{lee2025swap}. This attack, also taking advantage of the effects of crosstalk, focuses on the availability of quantum computers and emphasizes how devastating a disruption can be, reducing user output accuracy by intentionally positioning qubits. Side-channel attacks on quantum controllers on a NISQ quantum computing platform have also led to quantum circuit reconstruction~\cite{erata2024quantum}. Furthermore, timing-based side-channel attacks have been used on quantum cloud-based services to identify an individual quantum computer that executed a circuit with 10 measurements~\cite{lu2024quantum}. All of these attacks highlight the need for added protections and security measures that cloud providers must have to ensure the confidentiality, integrity, and availability of quantum circuits.

\subsection{Privacy and Authenticity Concerns}
Privacy can be a great concern for quantum systems with malicious actors having the ability to extract circuit information of users from various attack vectors. There have also been work highlighting some secretive measures quantum platform providers use which raise authenticity concerns. Fingerprinting methods have been tested on the trustworthiness of providers, which are accused of switching computers that a user originally selects to save costs and increase efficiency on their platforms~\cite{wu2024detecting}. This study used a controlled test to detect fraudulent services using a comparison of user-side and device-side fingerprints to determine the authenticity of a given computer. 

\subsection{Classical-Quantum Interface Solutions}
There have been a few proposed solutions to mitigate risks to the classical-quantum interface. One security measure uses camouflaging on rapid SQF circuits to prevent reverse engineering and has shown to greatly reduce exposure~\cite{kumar2019toward}. Another method uses logic locking to prevent outside attackers from being able to analyze the structural behavior of a circuit~\cite{jabbari2022hardware}. In addition, there is research on using entropy-based measures to detect any threats to the integrity of quantum systems~\cite{chehade2025entropy}.  

\subsection{Multi-Tenant Solutions}
The final subsection discusses mitigation strategies aimed at securing multi-tenant quantum cloud environments. While there are some methods to increase security in the quantum cloud, many proposals are still theoretical, are not very practical to implement, or would conflict with the multi-tenancy framework companies have employed. An analysis explores the use of various security mechanisms, such as zero-trust architectures, privacy-enhancing technologies, various encryptions, and access control, with the intention of suggesting secure solutions to eliminate some of the security threats in the quantum cloud. It suggests encryption for data at rest and in transit and stringent access control to help mitigate these problems~\cite{mendoza2023enhancing}. One paper suggests the creation of an antivirus that can scan a user's circuits for malicious patterns to help detect adversaries~\cite{deshpande2022towards}. However, much more work remains in the development of encryption algorithms and other methods, including an antivirus, to reduce multi-tenant risks~\cite{hayat2024securing}.

\begin{figure}[t]
\centering
\includegraphics[width=0.48\textwidth]{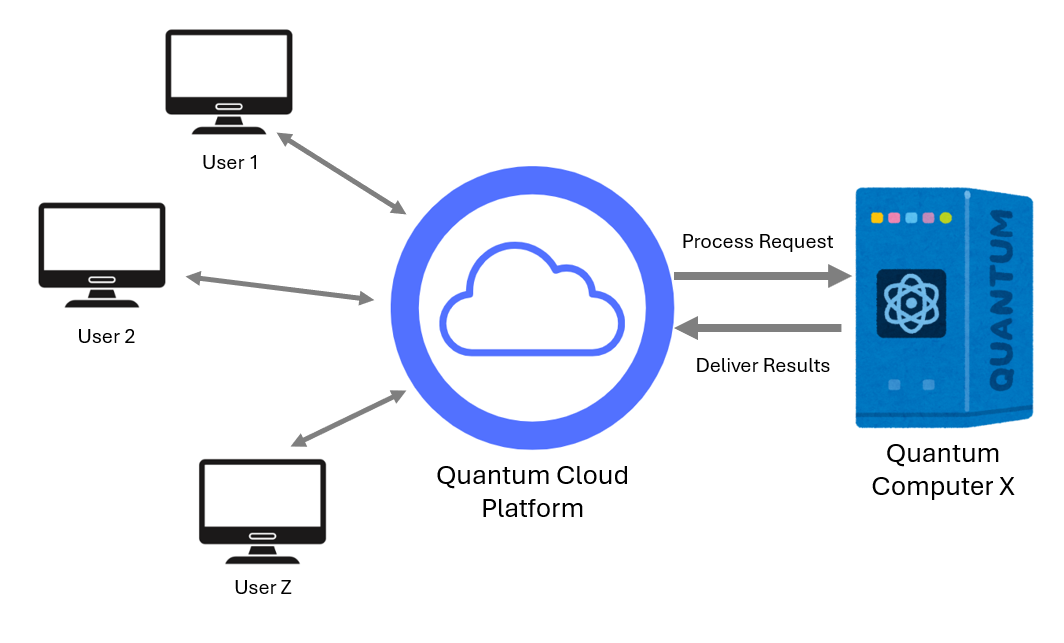}
\caption{Model of a Multi-Tenant Quantum Computing Platform on the Cloud}
\label{fig:Multi-Tenant Cloud}
\end{figure}

\section{Quantum Architectures}
Quantum computers are constantly developing in this new age of progress, and various companies are competing to be the leader of this new movement. Microsoft, for instance, announced that it has developed a type of quantum chip that can help them reach one million qubits~\cite{bolgar2025microsoft's}. With new advances happening this often, it is important to understand the background and overall design of quantum computers and how they process data. This section will provide an overview of the architecture of quantum computers and the different quantum computing platforms provided by IBM, Google, and others. It will cover the underlying hardware technologies and a high-level description of how these systems operate. It will also provide the rationale for why companies deploy could platforms using a multi-tenant framework. In addition, quantum workloads will be described in terms of how they are managed, scheduled, and executed in a cloud environment. The figure~\ref{fig:Multi-Tenant Cloud} provides a general model for the quantum cloud platform and users interaction with it.

\subsection{Background}
Quantum computing platforms have been developed and refined by leading technology providers which include IBM, Google, Honeywell, IonQ, Rigetti, and others. Each of them utilize various hardware technologies to power their systems. Quantum computing leverages the principles of quantum mechanics, such as superposition and entanglement, to perform computations that are impossible for classical computers. IBM, Google, and Rigetti predominately rely on superconducting qubits, which operate at extremely low temperatures close to absolute zero and are sensitive to environmental noise~\cite{singh2024survey}. In contrast, companies such as IonQ and Honeywell use trapped ion technology, which manipulates charged atoms with electromagnetic fields to achieve potentially higher fidelity operations~\cite{singh2024survey}. Understanding the design and functionality of these quantum cloud systems can help many to grasp the security risks of running quantum workloads. For example, the physical requirements of quantum hardware can introduce unique vulnerabilities, especially in the multi-tenant cloud, such as unauthorized interference or data leakage~\cite{choudhury2024crosstalk}.

\subsection{Quantum Cloud Integration}
Quantum workloads in a cloud environment follow a specific process for execution. Users submit jobs through cloud-based quantum computing platforms, where they get queued, scheduled, and executed on the quantum hardware they choose~\cite{wu2024detecting}. These platforms manage workloads by queuing jobs based on priority, resource availability, and the device the user selected. Security in this multi-tenant setting is critical because data must remain confidential during transmission and execution, protected by protocols such as encryption and  authentication mechanisms~\cite{choudhury2024crosstalk}. However, adequate security in these multi-tenant systems is not yet present due to the many crosstalk timing-based exploits which can put user circuits at risk. Another element that platform providers face is the balance between performance and security. Efficient execution must be balanced with isolation and security of quantum hardware to prevent unauthorized access while still ensuring the performance and low costs to provide quantum computing services.

\subsection{Security Considerations in Architectures}
Security is a very important factor in usefulness in an architecture, whether it is classical or quantum. Without security measures, no one would be able to use computers for important tasks such as messaging, data storage, or researching unexplored areas. Classical computers have made great strides over the past few years in improving security, however, quantum computers will make many of these measures irrelevant once it is fully realized. Thus, it is vital to create new security measures for quantum computers. In multi-tenant quantum cloud environments, this feat comes with challenges due to the resource and hardware sharing users face when trying to perform research. The confidentiality of data is currently in question on these systems due to crosstalk exploitations and side-channel attacks. The integrity of data is also at risk from the ability to create noise or alter circuits by positioning qubits.~\cite{lee2025swap} Additionally, the availability of devices is a concern when denial-of-service (DoS) attacks on a quantum computer will affect all users who currently share the device. Therefore, quantum service providers must factor all of these security considerations into their architecture designs to develop safe platforms for quantum development.

\section{Performance vs. Security}
Quantum cloud platforms like IBM Quantum, Amazon Bracket, and Microsoft Azure Quantum utilize quantum cloud systems to boost accessibility, but this creates a tradeoff between performance and security, especially in a multi-tenant framework. Performance is often throttled by security measures such as advanced encryption schemes, isolation methods, or noise injection. One example, blind quantum computing, protects user data by hiding inputs and outputs from the server. Secure protocols like the one mentioned before introduce significant communication and resource overhead. This will impact performance, leading to slower executions times in modern NISQ systems, where counts of qubits and coherence times are limited~\cite{nguyen2024quantum}.

The performance-security gap widens due to vulnerabilities like crosstalk between qubits or denial-of-service (DOS) attacks from miscalibrated qubits, which can disrupt shared resources. To counteract these, providers employ frequent calibrations and robust error correction, but these measures reduce computational efficiency~\cite{golec2024quantum}. Blind quantum computing in multi-tenant systems further conflicts with scalability, adding latency to ensure privacy across users~\cite{fitzsimons2017private}. Much research aims to develop protocols that minimize these trade-offs, striving for a balance that maintains both security and performance in shared quantum cloud environments~\cite{nguyen2024quantum}.

\section{Security Threat Landscape}
There are many types of known threats that affect quantum cloud systems, from new vulnerabilities exposed to existing threats that come naturally with these systems. There are also many works reporting on these threats in various survey papers, as well as others. This section is meant to provide references to key papers that already cover several topics of known security issues, as well as to identify what vulnerabilities do not have high-level survey coverage.

\subsection{Previous Survey Coverage}
One paper covers many side-channel attacks that affect superconducting quantum computers. This work includes some single-tenant and multi-tenant threats ranging from power-based attacks, timing-based attacks, fault injections, and crosstalk exploits~\cite{choudhury2024survey}. There is also research on power-based side-channel vulnerabilities in quantum computer controllers, so this topic will not be focused on this paper~\cite{xu2023exploration}. Additionally, there is much coverage on specific quantum hardware issues and how machine learning (ML) can potentially be used to help mitigate these threats~\cite{karimi2021hardware}. The impact of ML on quantum security is heavily researched on with discussions relating to quantum defenses like adversarial training, data privacy, and formal verification methods in this paper~\cite{franco2024predominant}.

\subsection{Survey and Research Gaps}
One gap in the paper on side-channels in superconducting quantum computers~\cite{choudhury2024survey}, is that they did not focus in detail on multi-tenant treats specifically. These issues were not explained comprehensively and did not include some key new research in exploits. Another paper covers side-channel attacks that target the classical-quantum interface in quantum computers~\cite{mustafa2024side}. These threats have limited research coverage, but this paper did well to address many of the vulnerabilities they pose. However, it can be difficult to grasp at a higher level to understand how these threats affect the overall landscape of security threats.
These two topics will be the primary focus of this survey paper
and how these exploits impact the security threat landscape.

\section{Classical-Quantum Interface Threats}
One important subsystem of a quantum computer is the classical-quantum interface, which refers to the interface between isolated qubits and the classical control or readout technology that is used in operation~\cite{reilly2015engineering}. Many side-channels that target the classical-quantum interface in quantum computers exploit vulnerabilities in single flux quantum (SFQ) circuits. These circuits are pivotal for refrigerator control and readout due to their high switching frequencies and low energy consumption per switch~\cite{mustafa2024side}. The SFQ to DC converter, essential for SFQ interface with CMOS technologies, has been shown to be particularly susceptible to exhibiting significant side-channel leakage~\cite{mustafa2023side}. 

\subsection{Threat Models for Attacks}
The threat model for classical-quantum interface attacks typically involves an insider, who has special access to electronics preferably at room temperature and a greater understanding of the system architecture. With this knowledge, the malicious actor can analyze variations in the bias current of SQF chips to potentially decode internal signals through monitoring~\cite{mustafa2024side}. Multiple attacks exploit SFQ-to-DC converter leakage. One exploit decodes control signals for two-qubit (CZ) gates by analyzing the bias current of current generators with 25 converters switching simultaneously~\cite{mustafa2023side}. Another attack targets qubit state readout using a Josephson photomultiplier, where an SFQ pulse signals a logical ‘1’, which allows attackers to infer the Hamming weight of the qubit’s bit string through bias current measurements~\cite{howington2019interfacing}. With multi-tenant systems commonly used, the effect of a compromise or information leak increases with more users exposed per quantum system. It is also possible for reverse engineering attempts to occur on circuits by both outside and inside attackers, as discussed in some papers below.

\subsection{Proposed Solutions}
All the mentioned vulnerabilities emphasize the importance of improved security to protect scalable quantum systems. One security measure through camouflaging method has been proposed on rapid SQF circuits to prevent reverse engineering attempts and has shown to greatly reduce the risks of exposure. However, this comes with a performance cost required in delay overhead and power overhead~\cite{kumar2019toward}. A method called logic locking has also been proposed to prevent outside attackers from being able to analyze the structural behavior of a design even when a circuit is obtained~\cite{jabbari2022hardware}. The implementation of this method also requires a reduction in performance to ensure security with the logic locked OR gate used to perform this method requiring 20 percent overhead leading to a 3.6 percent overhead in total. There is also recent research as of this paper's writing on using entropy-based measures to detect any threats to the integrity of quantum systems~\cite{chehade2025entropy}. Ongoing research should continue and prioritize developing robust countermeasures, such as improved circuit designs, noise injection techniques, and pursue new methods to secure the classical-quantum interface. There should also be access control measures in place to restrict as many insiders as possible from having the ability to obtain confidential information.

\section{Multi-Tenancy Vulnerabilities}
Modern quantum computing platforms operate using multi-tenancy, allowing resource sharing among users to optimize usage and save costs. This framework introduces many new vulnerabilities that exploit shared access among several users who want to keep their information confidential. It is very important to acknowledge the security landscape associated with multi-tenancy for researchers and developers who want to create circuits in research fields such as medicine, space applications, secure communications, and many others~\cite{moller2017impact}. This section will explore the various security threats that appear in multi-tenant quantum cloud environments. It will cover crosstalk interference, timing-based side-channels, qubit flipping attacks, and the risks of data leakage. Each threat will be analyzed for its impact on the confidentiality, integrity, and availability of quantum resources.

\begin{figure}[t]
\centering
\includegraphics[width=0.4\textwidth]{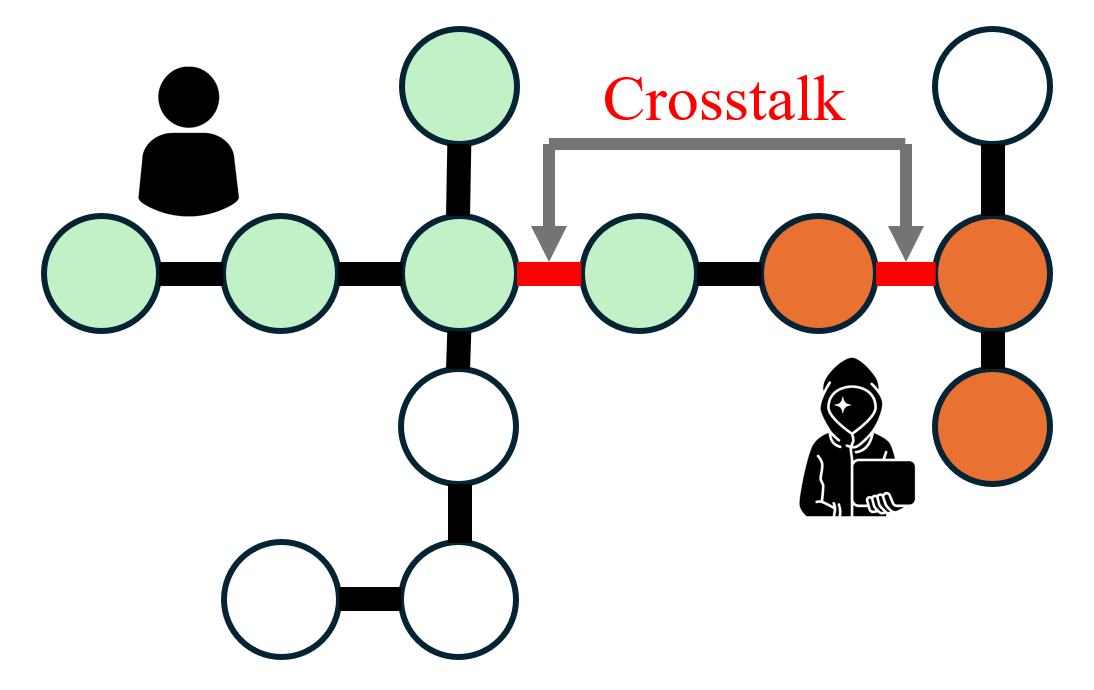}
\caption{Crosstalk-Channel Attack on a Quantum System}
\label{fig:Crosstalk Attack on Quantum System}
\end{figure}

\subsection{Crosstalk Exploits}
Crosstalk is the unwanted effect of interaction between sets of qubits in a quantum computing system. This has the effect of altering or leaking quantum information, thus affecting both integrity and confidentiality~\cite{ash2020analysis}. Figure~\ref{fig:Crosstalk Attack on Quantum System} showcases a basic crosstalk attack on a multi-tenant cloud system. One paper describes crosstalk in detail, as well as proposes a side-channel attack that uses it to extract unauthorized information on the victim's circuit by determining the number of CNOT gates in a quantum computer~\cite{choudhury2024crosstalk}. The attack framework introduced in this paper impacts the confidentiality and integrity of circuits of users sharing quantum hardware under a multi-tenant system. However, to perform the attack, the qubit numbers, total CNOT gates per qubit, time distribution of gates, and pairwise CNOT gate counts must be extracted to train their graph-based (GCN) model to accurately identify circuits. 

Another paper mentions that information can still be obtained from the effects of crosstalk in a multi-tenant NISQ system that goes under a reset~\cite{mi2022securing}. This side-channel attack occurs because a reset does not fully clear data like a complete system wipe, leaving the potential to acquire leftover information. The author's threat model requires the attacker to have control over the execution of a victim's program, use repeat measurements, and be co-located with the target victim. Through eavesdropping via crosstalk, information can be leaked across the reset gates on the same qubit. The paper advocates for the development of more secure resets that do not present this risk to confidentiality. It also acknowledges that full system wipes would solve the issue of data leakage, but it happens at a much slower speed.

\begin{table*}[t]
\centering
\setlength\tabcolsep{20pt}
\caption{Types of threats and their damage possibility stemming from the primary goals of the exploits on multi-tenant quantum cloud systems, focusing on their effects on the CIA Triad (confidentiality, integrity, and availability).}
\label{tab:Multi-Tenant Attack Ratings}
\begin{tabular}{@{} *{8}c @{}}
\toprule
Research Papers & Confidentiality & Integrity & Availability & Threat Level\\
\cmidrule(lr){1-5}
Crosstalk Side Channels~\cite{choudhury2024crosstalk}  & Yes & Yes & & High         \\
Active SWAP Attack~\cite{lee2025swap} & & Yes & Yes & Moderate       \\
Passive SWAP Attack~\cite{lee2025swap} & Yes & & & High       \\
Reset Operation Threats~\cite{mi2022securing} & Yes &  & & Moderate         \\
Reconstructing Circuits~\cite{bell2022reconstructing} & Yes &  & & Moderate         \\
Qubit Flipping Attacks~\cite{tan2025qubithammer} &  & Yes & & High       \\

\bottomrule
\end{tabular}
\end{table*}

\subsection{Timing-based Side-Channels}
Timing-based side-channel security exploits take advantage of the timing of computational processes to retrieve sensitive information such as part of a victim's circuit, keys, or passwords. Malicious actors can accomplish this through timing behavior analysis~\cite{choudhury2024survey}. These attacks can be dangerous due to their ability to be performed effectively on multi-tenant systems remotely, often in a secretive manner. Timing-based side-channel vulnerabilities have been explored on cloud-based quantum services working on both single-tenant and multi-tenant systems~\cite{lu2024quantum}. This research also demonstrates that it is possible to uniquely identify the quantum processor in use with just 10 measurements. The attack primarily targets the confidentiality of information, so it is important for quantum platforms to create mechanisms to keep data secretive.

Another side-channel attack involves timing through measurements before and after an execution of a circuit on IBM's cloud-based superconducting quantum computers. This exploit was able to achieve 60 percent accuracy to identify circuits on IBM's publicly available superconducting quantum computers~\cite{bell2022reconstructing}. The attack also exposes a risk to the confidentiality of user circuits on IBM's platforms and multi-tenant QCaaS platforms as a whole.

\subsection{Qubit Flipping Attacks}
There is new research as of writing this paper on qubit flipping attacks that can bypass existing security measures and leak important information. This paper proposes QubitHammer attacks which take advantage of qubit pulses to impact quantum circuits~\cite{tan2025qubithammer}. The paper presented four different attack situations as well as single and repeated attack pulse methods to perform these exploits. Their results conclude that current defense methods are not adequate enough to greatly reduce the risk of crosstalk. With this threat, these types of exploits can negatively affect the integrity of circuits by creating errors or disturbances. The research group claims attack success against existing security measures like dynamical decoupling, disabling qubit 0, crosstalk-aware qubit allocation, and active padding~\cite{tan2025qubithammer}. The primary goal of their work is to demonstrate the need for more effective and secure methods to prevent crosstalk and to stop these vulnerabilities inherent in multi-tenant quantum systems.

\subsection{Proposed Solutions}
There are several security solutions proposed to help mitigate risks in multi-tenant quantum cloud systems. One analysis delves into the use of zero-trust architectures, privacy-enhancing technologies, various encryptions, and access control, with the goal of recommending secure solutions to eliminate some of the security threats in the quantum cloud. It also suggests encryption for data at rest and in transit as well as strict access control to help mitigate these threats~\cite{mendoza2023enhancing}. Another proposal is for the creation of an antivirus that can scan a user's circuits for malicious patterns to help detect adversaries. This group did not actually create the antivirus, but instead made a theoretical one that could help reduce many threats in the quantum cloud~\cite{deshpande2022towards}. Other defensive measures (dynamical decoupling, disabling qubit 0, crosstalk-aware qubit allocation, and active padding) were also mentioned in the very recent qubit flipping attack, but were shown to be ineffective~\cite{tan2025qubithammer}. When using multi-tenant cloud systems, the risk of leakage from crosstalk has been clear and still have not had any viable solutions to stop this threat. Many works point to the need for more resources and time to be invested in new solutions to reduce multi-tenant risks~\cite{hayat2024securing,choudhury2024survey,tan2025qubithammer}.

\section{Security Threat Evaluations}
This portion of the paper reviews the general threats from the domains covered on the classical-quantum interface and multi-tenant systems and evaluates how concerning they should be for quantum platform providers and users. It will judge the overall damage these attacks can inflict, the feasibility of the attacks, and the areas these attacks primarily target. The ability of these attacks to perform against current defensive mechanisms will also be studied and incorporated into the overall risk the exploits pose. The evaluations will also transition into research gaps and challenges that need to be further studied and improved to limit these risks.

\subsection{Classical-Quantum Interface Risks}
The attacks present on the classical-quantum interface mainly focus on insider threats who have access to hardware and can make measurements that the majority of users could not. This threat presents less of a risk due to the lack of users who can perform this attack due to the knowledge and location required~\cite{mustafa2024side}. The exploits on this interface should not be discounted, however, since they can cause significant information leakage revealing internal signals used to recreate circuits without detection. The best countermeasure for this threat would be access control and least-privilege mechanisms to limit the amount of users who have the capabilities to perform this attack. Another method of promoting information confidentiality would be through camouflaging ~\cite{kumar2019toward} or logic locking ~\cite{jabbari2022hardware}, but both of these methods require overhead costs which are not attractive to cloud platform providers. This can lead to a fine line, where providers may not want to create too many secure measures that would reduce performance yet need to help ensure user circuits remain secret to keep their consumers. Overall, the attacks on the classical-quantum interface can deal moderate to significant damage depending on the goals of the adversary, but require a knowledgeable insider with great access to perform them thus making the attack less concerning than others such as multi-tenant threats. 

\subsection{Multi-Tenant Threats}
Multi-Tenant quantum cloud systems are vulnerable to numerous different exploits which many works have proposed especially in the past few years. This fact is very concerning considering that many quantum platform providers have adopted this system to increase performance and savings. While there have been many different defensive measures proposed by researchers, the crosstalk that is inherently a part of multi-tenant systems can still be used to gain partial information on user circuits~\cite{choudhury2024crosstalk}. This information is often significant enough to accurately reconstruct the circuits causing a breach in confidentiality of important new work in quantum computing, medicine, or finance~\cite{erata2024quantum}. There are also timing-based attacks, power-based attacks, SWAP attacks, and qubit flipping attacks, and others which all go down different avenues to either cause integrity damage to user circuits, by flipping qubits and creating errors, or confidentiality through revealing circuit information. Many of the main multi-tenant attacks are shown in table~\ref{tab:Multi-Tenant Attack Ratings} with their threat levels and impact on the CIA Triad. The most ideal solution would be for platform providers to use a different system than the current multi-tenant model that is exposed to these issues, but the costs make that unrealistic for them. Exploits in this area are very important to monitor and a major concern for both providers and users.

\section{Research Gaps and Challenges}
This section will identify the limitations of current security solutions and outline the unresolved challenges in securing multi-tenant quantum cloud environments. It will also emphasize the need for innovative security models. The roles of regulatory frameworks and the importance of cross-disciplinary collaboration in addressing these challenges will be mentioned.

There are several research gaps for new exploits and secure solutions in the quantum cloud. In terms of the classical-quantum interface, there needs to be more secure measures that have less overhead costs than the existing proposed methods of defense. Quantum platform providers are likely to not focus on security on this front if the cost is too high since it requires an insider with great access to perform. It can also be very difficult to defend an attack against someone with this level of knowledge and capabilities, so measures with significantly positive results will be necessary for these providers to embrace them. There is also limited research on exploits on the interface, so there requires much more research on how much damage can be caused through this attack vector as well as if it is possible for an attack to be done without needing an insider like the one mentioned~\cite{mustafa2024side}.

There is also a definite need for more research on secure measures to prevent crosstalk exploitation in multi-tenant systems. The new attacks being proposed that utilize crosstalk are able to bypass many existing methods, so new research is needed for circuits on these systems to remain confidential. The qubit flipping attack, QubitHammer, needs to be addressed in particular as it was able to introduce errors and disruptions into quantum systems in multiple different ways against many of the known defenses~\cite{tan2025qubithammer}. Finally, there is also a lack of survey papers that cover the threat landscape that exists in quantum cloud systems that this work aims to contribute to.

\section{Conclusion}
In this paper, we provide a comprehensive survey on the types of threats posing quantum computing cloud platforms. We cover the background of how these platforms operate and the reasoning behind their resource sharing. We discuss previous works breaking down multi-tenancy and the modern QCaaS architecture. We also delve into the security threat landscape behind these modern platforms as well as current proposed solutions to these risks. We focus heavily on describing the threats and solutions for the classical-quantum interface and multi-tenant quantum systems, two topics with limited coverage. A ranking of these exploits affecting the confidentiality, integrity, and availability of users and their ability to access the quantum computers was also used. Finally, new research in proposing solutions to the vulnerabilities and side-channel is encouraged. This paper aims to spread awareness of the ongoing threats that quantum platforms can face to researchers and developers who want to create innovative applications in a secure environment. Finally, we strive to highlight the need for future research and development of secure access control, encryption algorithms, and least privilege mechanics to address the limitations of these threats. With progress in this area, these gaps can be further addressed to create a secure future for quantum computing.

\balance

\bibliographystyle{IEEEtran}
\bibliography{main}

\end{document}